\documentclass{an}
\usepackage{graphicx}
\usepackage{times}
\usepackage{fancyhdr}
\usepackage{natbib}
\newcommand{\upp}{2}
\newcommand{\down}{1}

\newcommand{\citeq}[1]{(\ref{#1})}

\newcommand{\Hz}{\mathrm{Hz}}

\bibpunct{(}{)}{;}{}{}{,}

\begin{document}

%
%
%
%
%

\Issue{9}		
\Volume{326}            
\Yearsubmission{2005}   
\Yearpublication{2005}  
\Pagespan{864}{866}         
\DOI{10428}		

\headnote{Astron. Nachr. {\bf /} AN {\bf \volume}, No. \issue, \pages\
(\yearofpublic) {\bf / DOI} 10.1002/asna.\yearofsubmission\doi}

%
\fancyhead{}
\fancyhead[LE,RO]{\small\thepage}
\fancyhead[RE]{\small Astron. Nachr. {\bf /} AN~{\bf \volume}, 
No. \issue\ (\yearofpublic) {\bf /} www.an-journal.org}
\fancyhead[LO]{M. A. Abramowicz et al.: The correlations and anticorrelations in QPO data}

%
\fancyfoot{}
\fancyfoot[LE,RO]{%
\scriptsize%
\copyright\/\yearofpublic\ WILEY-VCH Verlag GmbH \& Co. KGaA, Weinheim}

\title{The correlations and anticorrelations in QPO data}

\author{
M. A. Abramowicz\inst{1}, D. Barret\inst{2}, M. Bursa\inst{3}, 
J. Hor\'{a}k\inst{3}, W. Klu\'{z}niak\inst{4}, P. Rebusco\inst{5}, 
G. T\"{o}r\"{o}k\inst{6}
}

\institute{
Department of Physics, G\"{o}teborg University, S~412~96 G\"{o}teborg, Sweden
\and
Centre d'Etude Spatiale des Rayonnements, CNRS/UPS, 9 Avenue du Colonel Roche, 31028 Tolouse Cedex 04, France
\and
Astronomical Institute, Academy of Sciences, Bo\v{c}n\'{\i} II 1401, 141\,31~Praha 4, Czechia
\and
Max-Planck-Institute f\"{u}r Astrophysik, D-85741 Garching, Germany
\and
Zielona G\'{o}ra University, Lubuska 2, 65-265 Zielona G\'{o}ra, Poland
\and
Institute of Physics, Silesian University in Opava, Bezru\v{c}ovo n\'{a}m.\ 13, 746\,01 Opava, Czechia
}

\date{Received; accepted; published online}

\abstract{Double peak kHz QPO frequencies in neutron star sources varies in time by a factor of hundreds 
Hz while in microquasar sources the frequencies are fixed and located at the line 
$\nu_\upp = 1.5\nu_\down$ in the frequency-frequency plot. The crucial question in the theory of 
twin HFQPOs is whether or not those observed in neutron-star systems are essentially different
from those observed in black holes. In black hole systems the twin HFQPOs are known to be in a 
3:2 ratio for each source. At first sight, this seems not to be the case for neutron stars.
For each individual neutron star, the upper and lower kHz QPO frequencies,
$\nu_\upp$ and $\nu_\down$, are linearly correlated, $\nu_\upp=A\,\nu_\down + B$, with the slope 
$A<1.5$, i.e., the frequencies definitely are not in a 1.5 ratio. In this contribution we show 
that when considered jointly on a frequency-frequency plot, the data for the twin kHz QPO 
frequencies in several (as opposed to one) neutron stars uniquely pick out a certain preferred 
frequency ratio that is equal to 1.5 for the six sources examined so far.
}

\correspondence{horak@astro.cas.cz}

\maketitle

\section{Introduction}
It has been recognized some time ago \citep{Bursa02} that the pairs of high-frequency 
QPOs observed in \textit{all} neutron star sources can be fitted by a single linear relation 
\begin{equation}
  \nu_\upp= 0.89\,\nu_\down + 375\,\mathrm{Hz} \;,
  \label{eq:common-fit}
\end{equation}
where $\nu_\upp$ and  $\nu_\down$ refers to the upper and lower observed kHz QPO frequency. The linear correlation between the pair frequencies becomes even more apparent if we restrict ourselves to individual sources, though the coefficients of the relation \citeq{eq:common-fit} slightly differ from source to source. A particular example is the case of \mbox{Sco~X-1}, where the observed frequencies are well fitted with the linear dependence \mbox{$\nu_\upp=(3/4)\,\nu_\down+450\,\Hz$} with the accuracy better than one percent.

\citet{abr03} pointed out that the  distribution of
frequency ratios is strongly peaked near the 1.5 value, but made 
no statement about the actual correlation of frequencies.
A concern has been raised, whether their result is a statistical
illusion \citep{BeloniMendezHoman05}. 

In this contribution we examine the linear
frequency--frequency correlations for six neutron-star sources: \mbox{4U~1820$-$30}, \mbox{4U~1728$-$34}, 
\mbox{4U~0614$+$09}, \mbox{4U~1608$-$52}, \mbox{4U~1735$-$44} and \mbox{4U~1636$-$53}, and
show that within errors the correlations are consistent with the
following statement: the plots of the six \mbox{$\nu_\upp=A\nu_\down+B$} relations
intersect in one point $[N_\down,N_\upp]$, 
with \mbox{$N_\down\!\approx\!600\,\Hz$}, and the ratio of the frequencies in 
the intersection point is \mbox{$N_\upp/N_\down\!=\!1.5$} to high accuracy. 
To demonstrate this, we plot the five pairs of coefficients, $A$ and $B$, 
and show that they are linearly anticorrelated.

\section{Anticorrelation between the slope and shift}

\begin{figure}[t]
\resizebox{\hsize}{!}{
\includegraphics[width=0.49\textwidth]{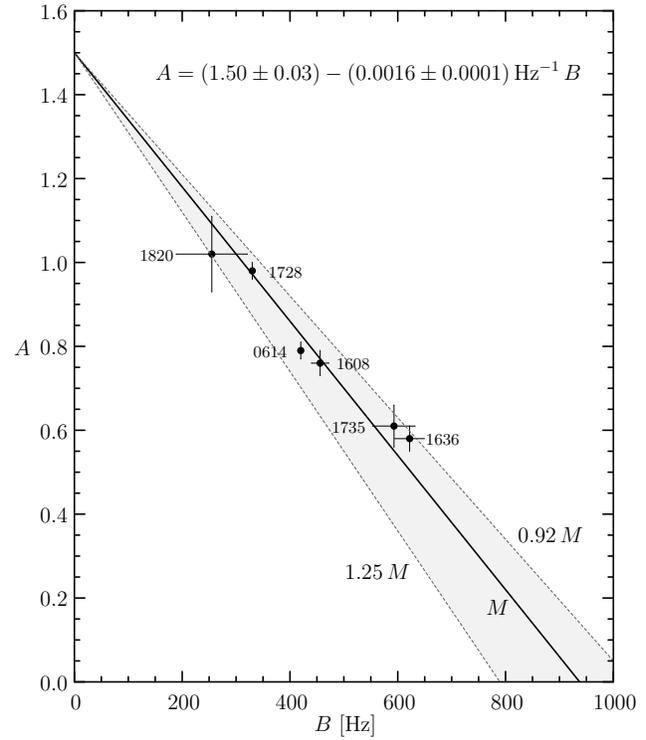}
}
\caption{The anticorrelation between the shift and slope and the effect of $1/M$-scaling. The figure shows the result obtained from the analysis of the data. Each point in the plot corresponds to the particular source. The shift $A$ and the slope $B$ correspond to the best linear fit of the $\nu_\down$--$\nu_\upp$ correlation. Clearly, $A$ and $B$ are anticorrelated among the sources with the best fit ({\it solid line}) given by equation \citeq{eq:anticorr}. In the spirit of $1/M$ scaling, it corresponds to a `typical' neutron-star mass between the examined system. Dashed lines represent the best fit recalled to $0.92 M$ and $1.25 M$. The value of the fit crosses vertical axis very close 1.5, which is in agreement with the idea that the excitation mechanism of QPOs is the 3:2 resonance.}
\label{fig:anticorr}
\end{figure}

We examine the six individual neutron star sources by fitting each of them with a linear formula
\begin{equation}
  \nu_\upp= A\,\nu_\down + B \;,
\end{equation}
where the coefficients $A$ and $B$ are referred to as the slope and shift, respectively. 
The resulting values of the slope and shift and of corresponding errors for each source are 
summarized in Table~\ref{tab:fits} and plotted in Figure~\ref{fig:anticorr} showing the 
slope--shift plane. The dependence \mbox{$A\!=\!A(B)$} strongly 
suggests that the two quantities are anticorrelated. The linear fit for the anticorrelation gives 
\begin{equation}
  A = (1.50\pm0.03) - (0.0016\pm0.0001)\,\Hz^{-1}\,B\;.
  \label{eq:anticorr}
\end{equation}
As it was mentioned above, this result is consistent with the statement that 
the six linear plots of the \mbox{$\nu_\upp$--$\nu_\down$} relations intersect 
in one point \mbox{$[N_\upp,N_\down]$}. The frequencies $N_\upp$ and $N_\down$ can
 be determined from the coefficients of the anticorrelation
\citeq{eq:anticorr}, from which we obtain
\begin{equation}
  N_\upp=(940\pm80)\,\Hz\,,\quad
  N_\down=(625\pm40)\,\Hz\,.
  \label{eq:point}
\end{equation}
The ratio of frequencies $N_\upp/N_\down$ at the intersection point 
equals to 1.5 with the accuracy of 15\%. Hence, it is remarkable that neutron stars
QPOs pick up the same 3:2 ratio common for the black-hole sources.
We can rephrase the equation \citeq{eq:anticorr} to the form
\begin{equation}
  A = \frac{3}{2} -\frac{B}{625\,\Hz}\;.
\end{equation}

\begin{table}[t]
\renewcommand{\arraystretch}{1.2}
\begin{center}
\caption[]{Best linear fits and their errors for the frequency--frequency correlation.}
\label{tab:fits}
\begin{tabular}{lcccc}
\hline
Source     &  $A$&$\Delta A$& $B$\,[Hz]&$\Delta B$\,[Hz]\\
\hline
4U~1636    & 0.58&0.03&622&27\\
4U~1608    & 0.76&0.03&456&16\\
4U~1820    &1.02&0.09&255&66\\
4U~1735    &0.61&0.05&593&39\\
4U~0614    &0.79&0.02&420&5\\
4U~1728    &0.98&0.02&330&5\\
\hline
\end{tabular}
\end{center}
\end{table}

The data points of Figure 1 have been obtained through a shift-and-add
technique, as described in \citet{Barret05} applied to the 
whole archival RXTE data  
available to date. The technique relies firstly on estimating the QPO  
frequency down to the shortest timescales permitted by the data  
statitistics, and secondly on identifying the QPOs for which the  
shift-and-add can be applied (generally the lower QPOs). Having a set  
of frequencies, the frequencies are then binned (typically with a bin  
of 10 to 20 Hz), and the individual PDS shifted to the mean frequency  
within the bin. The averaged PDS so-obtained is then searched for  
QPOs, and when both QPOs are present, their frequencies are obtained  
with a Lorentzian fit. A linear fit is then applied to the lower and  
upper QPO frequencies so obtained. This method has been applied to 4  
sources shown in Figure~1 (4U1820; 4U1608, 4U1735, 4U1636), whereas  
for 0614 and 1728, the linear fit has been applied to the frequencies  
measured with a shift and add technique applied to a continuous  
segment of observation (of typical duration close to the orbital  
period of the RXTE spacecraft). Both methods should yield consistent  
results: the smaller error bars the points of 0614 and 1728 can be  
explained by a larger number of frequencies involved in the linear  
fit. A more careful comparison of the results of the two methods  
(with estimates of the potential biases, such as only the narrower  
QPOs are detected, hence with small errors on the frequency  
determination)      is clearly required and will be part of a  
forthcoming paper (Barret et al. 2005b in preparation).

\section{A possible explanation}
It has been suggested that QPO arises from a resonant interaction between the radial and vertical oscillation modes in 
relativistic accretion flows. The strongest possible resonance occurs at the radius, where the radial and vertical 
epicyclic frequencies are in the 3:2 ratio. 

In general, the frequency and the amplitude of non-linear oscillations are not independent. In the lowest order of approximation
the observed frequencies $\nu$ differ from the eigenfrequencies $\nu_0$ of oscillators by corrections proportional to 
the squared amplitudes. We consider system having two oscillation modes, whose eigenfrequencies are 
$\nu_{\down,0}$ and $\nu_{\upp,0}$. The frequencies of non-linear oscillations may be written in the form
\begin{equation}
  \nu_\down = \nu_{\down,0} + \Delta\nu_\down\,, \quad
  \nu_\upp = \nu_{\upp,0} + \Delta\nu_\upp\,.
  \label{eq:corrections}
\end{equation}
The frequency corrections $\Delta\nu_\down$ and $\Delta\nu_\upp$ are proportional to the squared amplitudes
\begin{equation}
  \Delta\nu_\down=\kappa_\down a_\down^2 + \kappa_\upp a_\upp^2\,,
  \quad
  \Delta\nu_\upp=\lambda_\down a_\down^2 + \lambda_\upp a_\upp^2\,,
  \label{eq:correct2}
\end{equation}
where $\kappa_\down$, $\kappa_\upp$, $\lambda_\down$ and $\lambda_\upp$ are constants depending on non-linearities
in the system.

Let us suppose that for some reason the two amplitudes $a_\down$ and $a_\upp$ are correlated. Thus, 
one may consider the amplitudes as functions of a single parameter $s$,
\begin{equation}
  a_\down=a_\down(s)\,, \quad a_\upp=a_\upp(s)\,.
  \label{eq:aacorr}
\end{equation}
Such relations can be considered as a natural consequence of an interplay between the resonance excitation mechanism 
and the dissipation of the energy in the system. 

It follows that the frequencies of non-linear oscillations $\nu_\down$ and $\nu_\upp$ are correlated as well. 
Up to the linear order in $s$, we obtain from equation \citeq{eq:corrections}
\begin{eqnarray}
  \nu_\down &=& \nu_{\down,0} + \nu_{\down,0} F_\down s + \mathcal{O}(s^2)\,,
  \label{eq:obsomr}\\
  \nu_\upp &=& \nu_{\upp,0} + \nu_{\upp,0} F_\upp s + \mathcal{O}(s^2)\,,
  \label{eq:obsomth}
\end{eqnarray}
where the coefficients $F_\down$ and $F_\upp$ are given in terms of constants 
$\kappa_\down$, $\kappa_\upp$, $\lambda_\down$ and 
$\lambda_\upp$ of frequency corrections (\ref{eq:correct2}) and of the derivatives 
$\mathrm{d}a_\down/\mathrm{d}s$ and
$\mathrm{d}a_\upp/\mathrm{d}s$ at the point $s=0$.

Isolating the parameter $s$ from equations (\ref{eq:obsomr}) and (\ref{eq:obsomth}) we get the linear correlation
between the observed frequencies
\begin{equation}
  \nu_\upp= A\nu_\down + B\,,
  \label{eq:ffcorr}
\end{equation}
where the slope and shift are respectively given as
\begin{eqnarray}
  A&=&\frac{\nu_{\upp,0}}{\nu_{\down,0}}\,Q\,,
  \label{eq:slope}
  \\
  B&=&\nu_{\upp,0}(1-Q)
  \label{eq:shift}
\end{eqnarray}
and we define $Q$ as $Q\equiv F_\upp/F_\down$.

The observed frequencies $\nu_1$ and $\nu_2$ of systems with different amplitude prescription (\ref{eq:aacorr}) 
are correlated in a different way. Any particular value of $Q$ leads to particular values of the slope and the shift.
However, if the eigenfrequencies of the systems are similar, the slope and shift are necessarily anticorrelated. 
Solving the equations (\ref{eq:slope}) and (\ref{eq:shift}) for the parameter $Q$, we obtain 
\begin{equation}
  A = \frac{\nu_{\upp,0}}{\nu_{\down,0}}-\frac{1}{\nu_{\down,0}}\,B\,.
  \label{eq:anticorr3}
\end{equation}
From the resonance condition it follows that the eigenfrequency ratio $\nu_{\upp,0}/\nu_{\down,0}$ is approximately 3/2.
Therefore we arrive at
\begin{equation}
  A = \frac{3}{2} -\frac{1}{\nu_{\down,0}}\,B \,.
\end{equation}

\section{Discussion and Conclusions}
In black-hole sources the observed QPO frequencies are fixed and always have the ratio 3/2. 
It has been recognized that their actual frequencies scales inversely with mass $M$ 
assuming a similar value of the spin (\citealt{McClintockRemillard05}, 
[van~der~Klis], [T\"{o}r\"{o}k]\footnote{Articles by other authors in this Volume.}). In neutron-star 
sources the frequencies are not fixed, but their distribution 
seems to cluster around a single line for each individual source. By linear fitting the observed data, 
we have found out that these lines intersect around a single point \mbox{$[N_\down,N_\upp]$}, which 
have coordinates given by equation \citeq{eq:point}. The fact that 
the frequencies are close to 3:2 ratio supports the idea that there is
a similar mechanism at work in both classes of sources. Moreover, if we extend the black-hole $1/M$ 
scaling law up to the frequency of the intersection point, we obtain a mass of order of one solar mass
(assuming zero angular momentum for neutron stars), which provides an additional hint.

Assuming that the $1/M$-scaling can be adopted also to the neutron-star QPOs, 
the fact that the individual positions of sources in the $A$--$B$ plane do not strictly 
follow the anticorrelation line can be attributed to small differences in neutron-star masses. 
By scaling the 614~Hz frequency of equation $\citeq{eq:anticorr}$, we find that the $A$--$B$ is steeper
or softer for more massive or less massive sources, respectively. This is demonstrated by the shaded region in Figure~\ref{fig:anticorr}. Under this assumption the deviation in the masses of examined neutron stars should not be greater than $\sim 20\%$.

\acknowledgements
It is a pleasure to acknowledge the hospitality of Sir Franciszek Oborski, 
the master of the Wojnowice Castle, where a large part of this work was done. 


\end{document}